\documentstyle[12pt,epsf]{article}
\newcommand{\ds }{\displaystyle}

\newcommand{\be}{\begin{equation}}
\newcommand{\ee}{\end{equation}}
\newcommand{\bea}{\begin{eqnarray}}
\newcommand{\eea}{\end{eqnarray}}
\newcommand{\ci}{\cite}
\newcommand{\bi}{\bibitem}

\newcommand{\dd}{\partial}

\newcommand{{\bfna}}{\mbox{\boldmath$\vec{\nabla}$}}

\newcommand{\rr}{\vec{\bf{r}}}

\def\dal{\,\lower0.3ex\vbox{\hrule\hbox{\vrule\kern2pt\vbox{\kern4pt\kern4pt}
\kern2pt\vrule}\hrule}\,}

\begin{document}

\title{\sl A Helium nanodrop bouncing off a wall}
\vspace{1 true cm}
\author{G. K\"albermann$^*$
\\Soil and Water dept., Faculty of
Agriculture, Rehovot 76100, Israel}
\maketitle

\vspace{3 true cm}
\begin{abstract}

We investigate numerically the quantum 
collision between a stable Helium nanodrop
and an infinitely hard wall in one dimension.
The scattering outcome is compared to the same event omitting
the quantum pressure. 
Only the quantum process reflects the effect of diffraction of wave packets in
space and time.

\end{abstract}

$^*${\sl e-mail address: hope@vms.huji.ac.il}

\newpage

\section{\sl Introduction}

In recent years we have described the effect named: {\it Diffraction of
wave packets in space and time}. The effect consists in a multiple
peak wave train that is generated in a scattering event, by the interference
between the incoming and the scattered packets.
The wave train lasts to infinite time, due to the spreading of
the incoming packet, that catches up with the scattered packet.\ci{k1}-\ci{k6}

The effect occurs in matter wave scattering. It persists only
for narrow enough packets compared to the scatterer extent.
The effect was demonstrated numerically and analytically in various
settings. 
As a possible laboratory experiment to verify
 the effect experimentally we suggested a collision between a 
Helium drop and a reflecting wall.\ci{k3}
In ref.(\ci{k3}) we omitted completely the self-interaction of the drop.
The impinging packet that represented statistically the behavior
of the drop was taken as a Gaussian packet, for which we could provide
exact analytical expressions.
However, the self-interaction provides the
binding energy of the drop and hinders spreading, that is
crucial for the appearance of the diffraction pattern. 
A more realistic calculation was needed.

Helium is relatively a weakly bound liquid,
 even close to zero degrees Kelvin. Weaker than liquid water or
other liquids at ambient temperature.
The binding energy per Helium$^4$ atom in the superfluid phase is
around $\ds E_b=-7.5 ^0K$\ci{strin1} whereas for liquid water it is
 around $\ds -5500 ^0K$, depending on temperature and pressure.

The Helium atoms are neutral and do not benefit from strong Hydrogen-like 
 bonds for their binding. The apparent weakness of
the interactions is nevertheless misleading. A Helium drop
interacts with itself in a much stronger manner than 
a Bose-Einstein condensate (BEC)\ci{dalfovo}, and even there
the nonlinear effects of the self-interaction are important.
In particular the interatomic forces dominate over the 
so called 'quantum pressure' or quantum potential in the bulk of
the drop. Inside the drop the density is constant and the
quantum pressure vanishes. Near the edge of a drop, curvature effects
take over and the quantum pressure becomes relevant.

In the present work we investigate numerically the quantum scattering 
of a Helium droplet bouncing off an infinitely hard wall.
The aim is twofold: Firstly an investigation like the present has never been
 carried out before and it is interesting in its own merit and,
secondly we would like to have a more realistic treatment of
the diffraction of wave packets effect in the context of liquid Helium.

In order to simplify the treatment as much as possible we resort
to the {\it density functional} approach.

The density functional phenomenological method for interacting 
quantum systems, is based upon the solution of the one-body Schr\"odinger 
equation carrying self-interaction nonlinear terms that depend
 on the density only.
Many topics are treated nowadays by means of the
density function theory. To name a few; 
electron transport in solids\ci{weinberger}, 
electronic excitations\ci{onida}, soft condensed matter\ci{likos}  
phase transitions in liquid crystals\ci{singh}, phonons in solids \ci{fritsch,
baroni}, colloids\ci{lowen}, liquids and nuclei \ci{ ghosh}, atoms and molecules
 \ci{nagy}, quantum dots\ci{reimann}, etc.
As a numerically viable phenomenological theory,
 is even becoming the dominant method for treating many body quantum
systems.
This is evidenced by the number of papers on the topic that, 
by the late 1990's, surpassed the works using the Hartree-Fock method.\ci{arg}

Density functional theories resort to the solution of a one-body nonlinear 
Schr\"odinger equation for a particle of mass {\it m}

\bea\label{schr}
i\frac{\dd\psi}{\dd t}~=~-\frac{1}{2~m}{\bfna}^2\psi~+[O(|\psi|^2)+~U(\rr)]\psi
\eea

where $\ds O(|\psi|^2)$ is a nonlinear and sometimes nonlocal\ci{strin1}
 functional of the density $\ds \rho= |\psi|^2$, and $\ds U(\rr)$ an
external potential.
The commonly used mean field equation for BEC systems belongs to this class 
of equations.\ci{dalfovo}
In this case it is referred to as the Gross-Pitaevskii equation.

Density functional theory has been extremely successful in reproducing the
properties of Helium nanodroplets.\ci{strin1}
Besides the mean binding energy per particle, average density, incompressibility
 and surface tension at zero temperature both for Helium$^4$ and Helium$^3$,
in a stationary state\ci{stringari}, 
 it accounts for capillary effects at low temperatures
\ci{calbi}, the phase diagram of liquid-vapor coexistence\ci{biben}, as well as 
 the excitation spectrum (phonon-maxon-roton) of liquid Helium\ci{strin1}.
The model has been verified experimentally in scattering reactions\ci{harms}.
It is currently being applied to other areas of superfluid dynamics such as
electron bubbles,\ci{eloranta} adsorption on plates of Alkali 
metals\ci{szybisz}, vortex line pinning\ci{barranco}, etc.

The density functional theory is therefore a respectable 
method for the treatment of Helium nanodroplets scattering.
In the next section we present results for the scattering of Helium nanodroplets
 in one dimension with and without the quantum potential.
The last section provides some comments emerging from the observation
 of the numerical data.

\newpage
\section{A helium drop colliding with a wall}

A nonlinear, local, self-interaction of the Helium drop was proposed some time
ago by Stringari and Treiner\ci{stringari}. 
The parameters of the density functional
were fitted to the known values of the binding energy per atom
in the bulk, the incompressibility, the infinite atom matter density
and the surface tension. Improvements to this functional
seem to require nonlocal terms\ci{dalfovo}. Such terms are engineered
 to reproduce the excitation spectrum of liquid Helium below
the $\lambda$ transition point in a more accurate manner.
In the present work we opt for the local version that is more manageable
computationally and still quite successful phenomenologically.

The operator $\ds O(\rho)$ of eq.(\ref{schr}) is given by

\bea\label{o}
O(\rho)=b\rho+\frac{c}{4}(4+2\gamma)\rho^{1+\gamma}-2~d~\bfna ^2\rho
\eea

With $\ds b=-888.81~^0K\AA,~c=1.04554~10^7~^0K~
\AA^{3+3\gamma},~\gamma=2.8,~d=2383^0K~\AA^5$.

For the one-dimensional case treated presently we use the same parameter set,
assuming independence from the {\it y,z} directions, i.e. an infinite
planar slab.

The stationary solutions of eq.(\ref{schr}) with the self-interaction of 
eq.(\ref{o}), in one spatial dimensions are obtained from the substitution
$\ds\Psi(x,t)=e^{i\mu~t}\phi(x)$, with $\ds\mu$, the chemical potential.
For finite size drops, $\ds \mu$ is a parameter , whose
value determines the number of particles in the drop and tends to the bulk
 binding energy when the number of particles tends to infinity.
For Helium$^4$, it is $\mu_{\infty}=-7.15 ^0K$.\footnote{
The transformation to standard energy units proceeds by multiplication
with Boltzmann's constant {\it k}}

Stationary solutions are found by integration of the equation\ci{stringari}

\bea\label{stat}
\mu=b~\rho+\frac{2+\gamma}{2}~c~\rho^{1+\gamma}-(2~d+\frac{1}{4~m\rho})
~\frac{d^2\rho}{dx^2}+\frac{\bigg(\frac{d \rho}{dx}\bigg)^2}
{8~m~\rho^2}
\eea

or equivalently

\bea\label{stat1}
\bigg(\frac{d\rho}{dx}\bigg)^2~=~\frac{8~m\rho}{8~m\rho~d+1}\bigg(
-\mu~\rho+\frac{b~\rho^2}{2}+\frac{c~\rho^{2+\gamma}}{2}\bigg)
\eea

The integration of eq.(\ref{stat1}), proceeds by a choice of $\ds \rho(x=0)<
\rho(x=0)_{\infty}~=~0.021836~\AA^{-3}$, the bulk helium liquid density
for atomic matter. Eq.(\ref{stat1}), with $\ds
\frac{d\rho}{dx}|_{x=0}~=~0$, fixes the value of the chemical potential
$\ds \mu>\mu_{\infty}$.
Numerical roundoff error limits the possibilities of approaching $\ds \rho(x=0)_
{\infty}$ closer than $\ds 10^{-7}$ and still obtain a profile density
that decays at infinite distances. We therefore chose the closest value we could
 and used this profile density in all the scattering events to be presented 
below. The parameter for this density profile are $\ds \rho(x=0)~=~0.02183599$,
that yields $\mu~=~-6.71 ^0K$.
The number of particles per unit area

\bea\label{np}
N~=~\int_{-\infty}^{\infty}~dx~\rho(x)
\eea

for this case is $N~=~1.288~\AA^{-2}$, or an effective length 
$\ds~X_{eff}=\frac{N}{\rho(x=0)}~=~59~\AA$. 
The drop then extends some $\ds 59\AA$ along the {\it x}
 axis and is infinite along the {\it y,z} plane.
The solution in the three dimensional case is not much
 different, in this case, the drop contains around 4500 atoms.
Figure 1 shows this initial density profile.
\begin{figure}
\epsffile{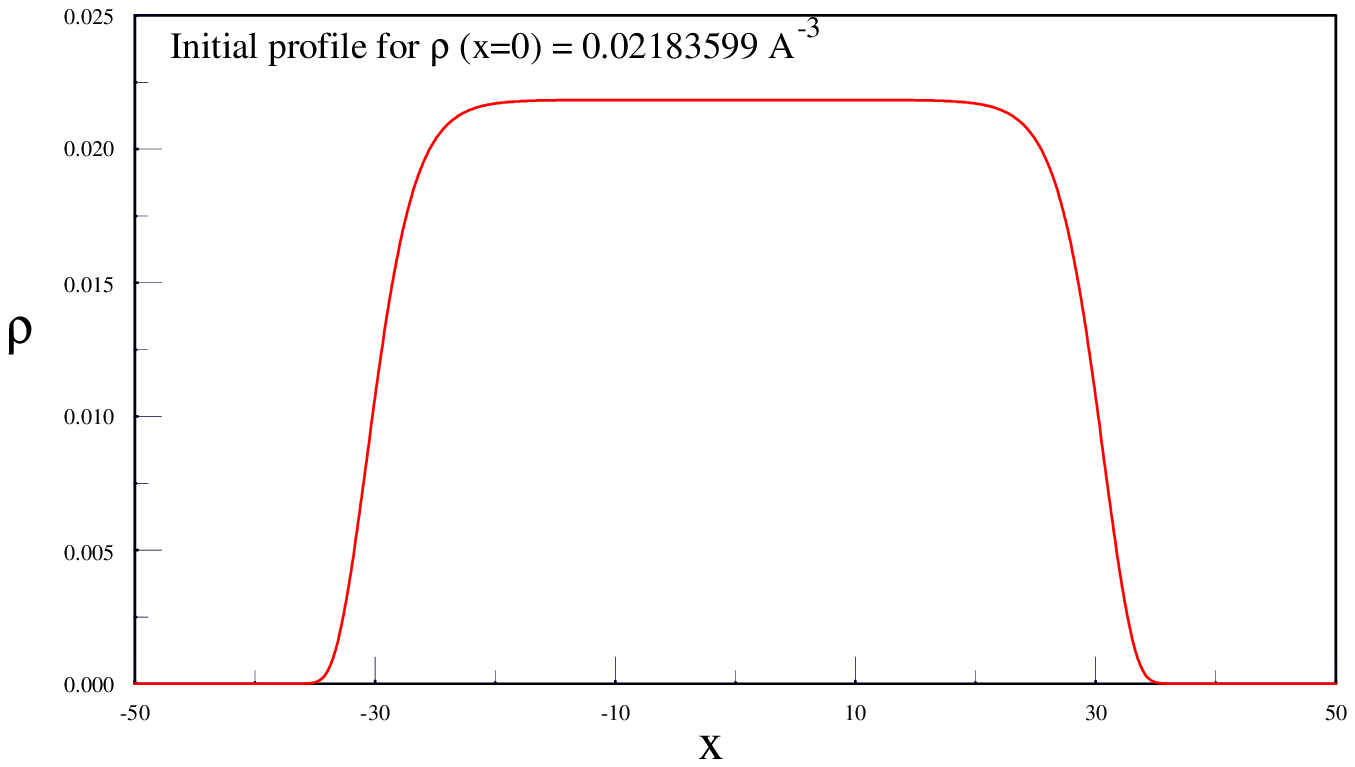}
\caption{\sl Density profile of a Helium drop
in $\AA^{-3}$ of for $\rho(x=0)~=~0.02183599~\AA^{-3}$ as a function
of distance in $\AA$}
\label{fig1}
\end{figure}

The details of the scattering are not very
sensitive to the number of particles and the extension of the drop.
The majority of the particles reside within the bulk. The diffraction effects
depend on the side wings of the distribution in figure 1 and are therefore
almost the same for any number of particles, because the width of these
wings is determined by the parameters of the nonlinear potential only.

In the next section we will show results for smaller size
drops whose profile resembles a Gaussian. For such drops the 
interference effect is cleaner, but not qualitatively different.

The scattering starts with a drop impinging from the left onto an
infinitely hard wall located to the far right. We took
a drop centered at $\ds x_0~=~110~\AA$ with velocity {\it v}, defined
by

\bea\label{psi0}
\psi(x,t=0)~=e^{i~m~v~(x-x_0)}\sqrt{\rho(x)}
\eea

The wall is located at $\ds x=150~\AA$. We use a five step predictor corrector
method\ci{eloranta} for the numerical integration. We demand an accuracy
of more than $\ds 0.1\%$ in both the conservation of normalization of the wave
and the energy. This strict demand requires a tiny time step of around
$\ds 10^{-17}$ sec, and consequently lengthy runs of around $\ds 10^7$ 
iterations. A larger time step causes rapid deterioration of the accuracy
and runaway behavior. A tiny time step is needed, 
due to the large values of the constants entering the self-interaction.

It was found that for packet velocities below around 
$\ds v~=~50\frac{m}{sec}$ the drop collides with the wall elastically
 with a barely noticeable distortion. Such a threshold corresponds
to a momentum transfer per particle at the wall of the order of
 $\ds \Delta p~=~2~m~v\approx 0.62 \AA^{-1}$. This is a typical value for
the excitation of ripplons, at the surface of a drop.\ci{ebner}

The drop is travelling without strictures, hence there is no superfluid
limiting velocity, it can remain superfluid 
up to any velocity, in practice the drop flows
over a surface that limits the maximum frictionless speed. 
The experiment may also be viewed as one
in which the wall is moved crashing into the stationary drop. 
We have performed such calculations
in order to check our numerical schemes, and found them to agree with
the results of a moving packet and a fixed wall as expected.

Figure 2 exemplifies the result of the scattering for various
velocities and times, such that $\ds \Delta x~=~v~t$, for all of them 
is the same in the free case. Above the threshold velocity there
appears a multiple peak structure that receds faster than the packet.
It resembles the diffraction in space and time structures seen in refs.\ci{k1}-
\ci{k6}. For high enough velocities, the collision is inelastic.
The energy is transferred into waves that recede faster than
the bulk of the drop. The drop is almost stalled at the wall. 
In the bottom graph of figure 2 the collision process 
is still underway and the multiple peak structures are still being generated.
There is a background hump under the peaks in the middle graph. This
elevation is an incoherent background, similar to the one occurring
in the {\it diffraction in space and time phenomenon}\ci{k1}-\ci{k6} for
wide packets. In the next section we will show pictures for a thinner
 drop, for which the background is almost absent.
\begin{figure}
\epsffile{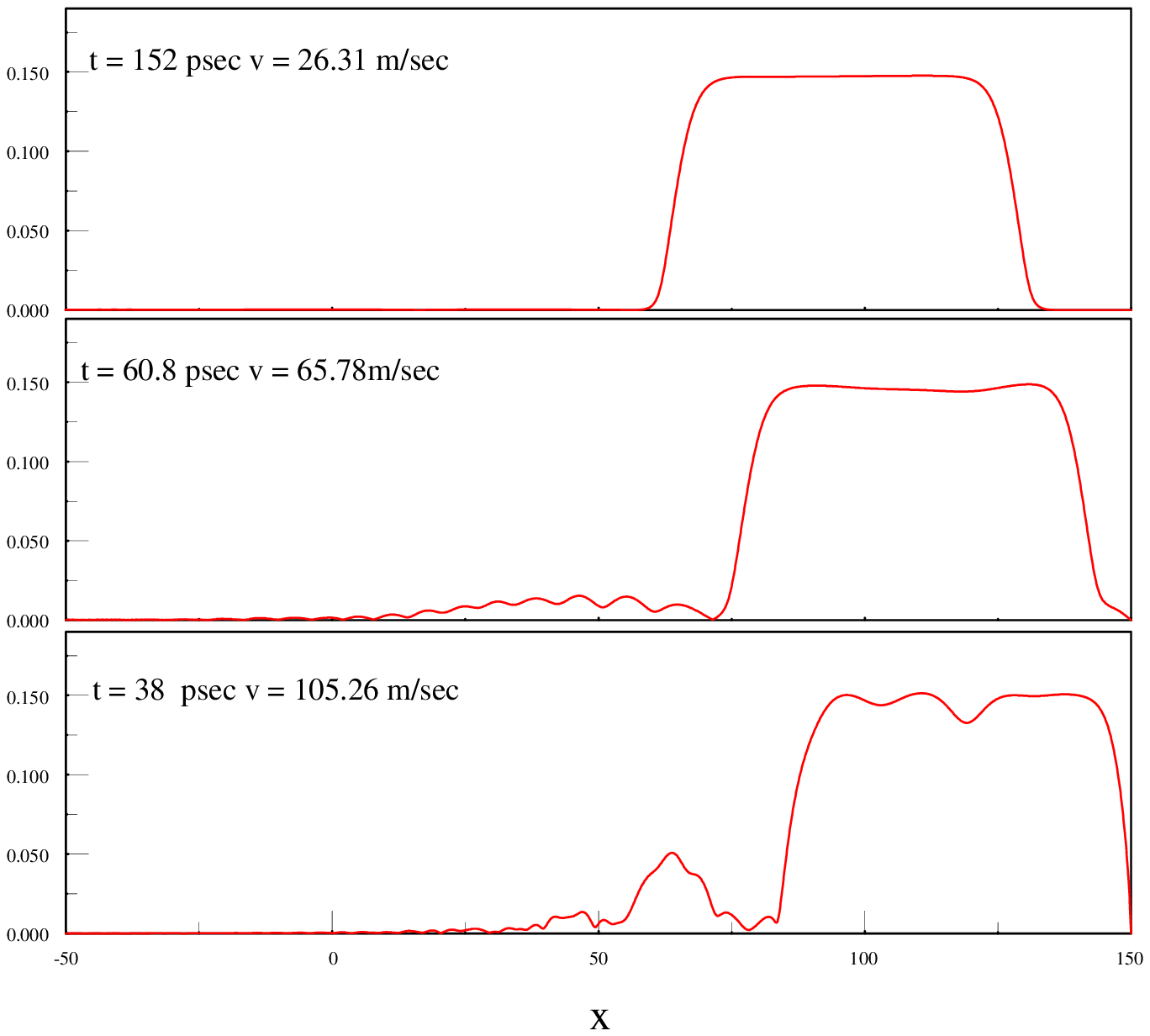}
\caption{\sl $|\Psi|$ as a function of distance
in units of $\AA$ for various velocities and times}
\label{fig2}
\end{figure}

\begin{figure}
\epsffile{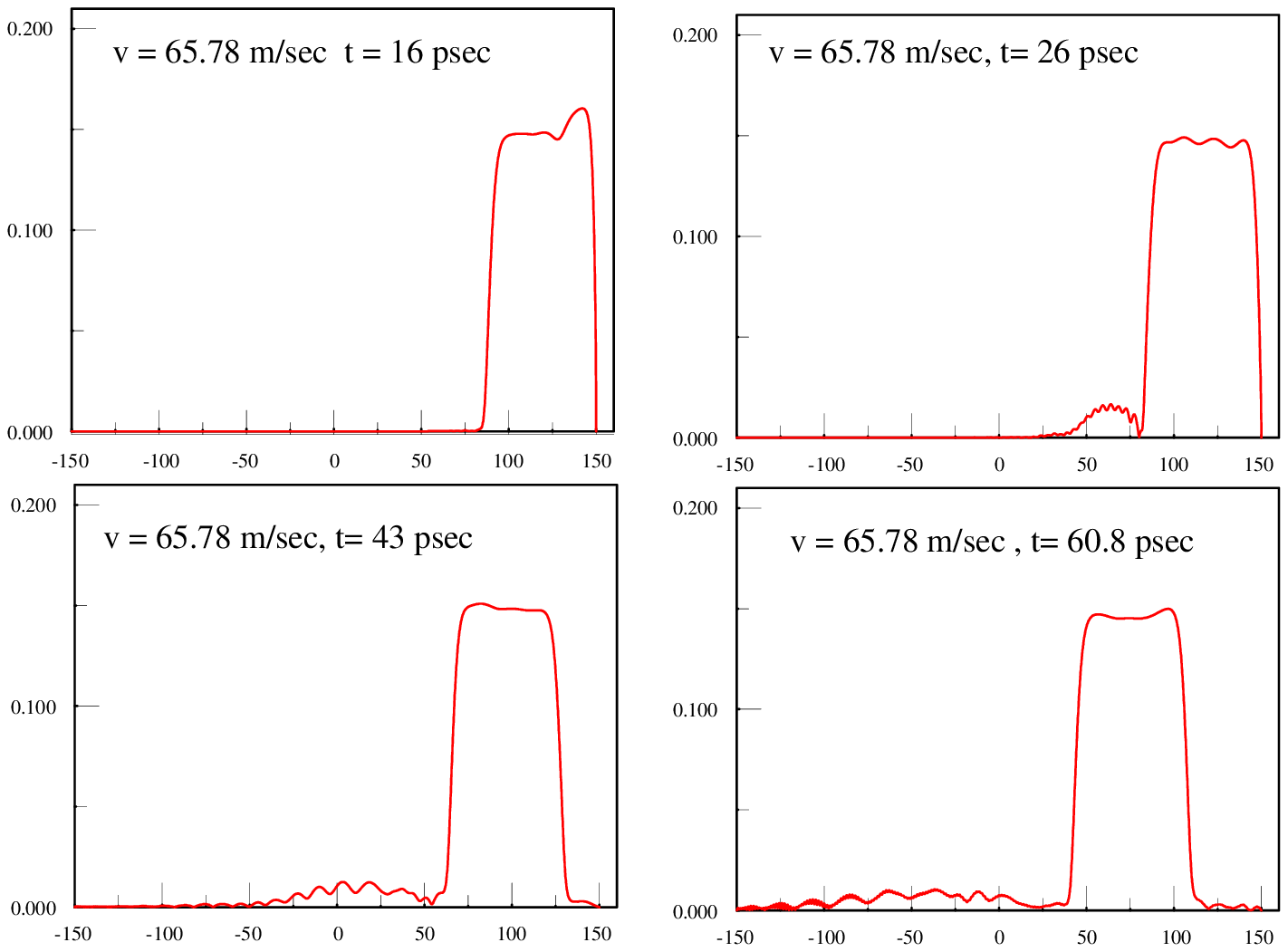}
\caption{\sl Time snapshots of $|\Psi|$ as a function of distance
in units of $\AA$ for $v~=~65.78\frac{m}{sec}$}
\label{fig3}
\end{figure}

Figure 3 shows the evolution of the scattering process for a 
velocity of $\ds v~=~65.78~\frac{m}{sec}$.\footnote{ The awkward value
of the velocity originates from a convenient choice of time units}

The density fluctuations inside the packet due to the collision, 
are progressively expelled out of the drop.
This multiple peak tail wave, is a product of 
both the incoming and the reflected waves interfering with each other.

In order to confirm the hypothesis that
 the interference effects are responsible for
the multiple peak structures, we performed parallel calculations for the
analog classical drop, by subtracting the quantum potential in the
Schr\"odinger equation.
In eq.(\ref{schr}) $O(\rho)$ is replaced by $\ds O(\rho)~-~U_{quantum}$, with
$\ds U_{quantum}~=~-\frac{1}{2~m~\sqrt{\rho}}\frac{d^2\sqrt{\rho}}{dx^2}$.
For the classical scattering case , we use the same initial profile 
as the one depicted in figure 1, although it is not
a stationary solution with the quantum potential subtracted. As a matter
of fact there are no stationary solutions without the quantum potential.
This may be easily seen by expanding around the vacuum.

Such an expansion produces the equation

\bea\label{asymp}
\frac{d^2\rho}{dx^2}~=~\frac{b}{2~d}\rho
\eea

whose solutions are oscillatory, with the density becoming eventually negative.

Figure 4 depicts the classical scattering pictures corresponding to the
quantum events of figure 2. As is
evident from the figure, the classical collision does not produce
a multiple peak tail. The collision results in 
large density fluctuations in the bulk and noisy fluctuations
at the edges.

The corresponding classical packet evolution for $\ds v~=~65.78~\frac{m}{sec}$
appears in figure 5.
\begin{figure}
\epsffile{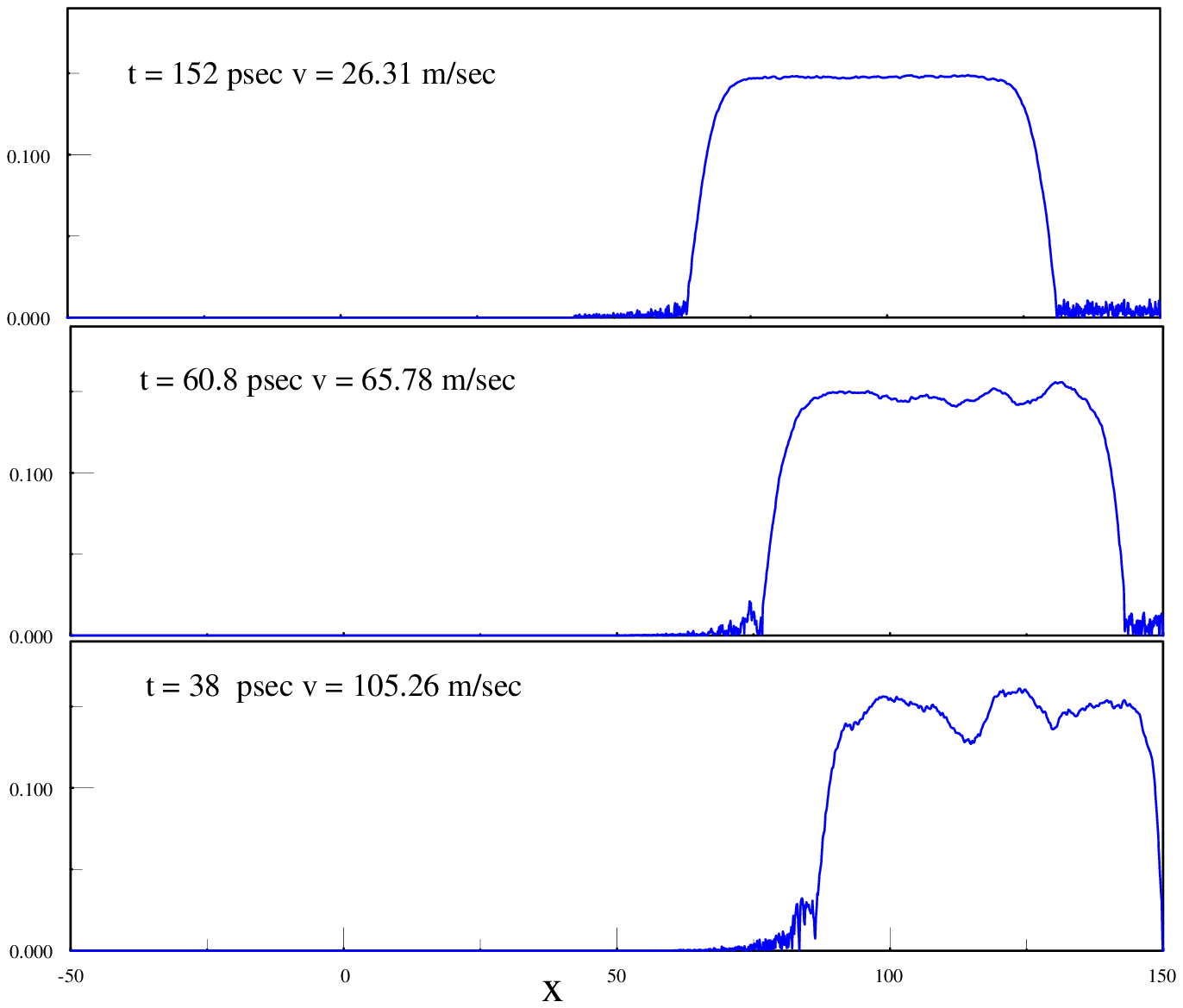}
\caption{\sl $|\Psi|$ as a function of distance
in units of $\AA$ for various velocities and times, classical case}
\label{fig4}
\end{figure}

\begin{figure}
\epsffile{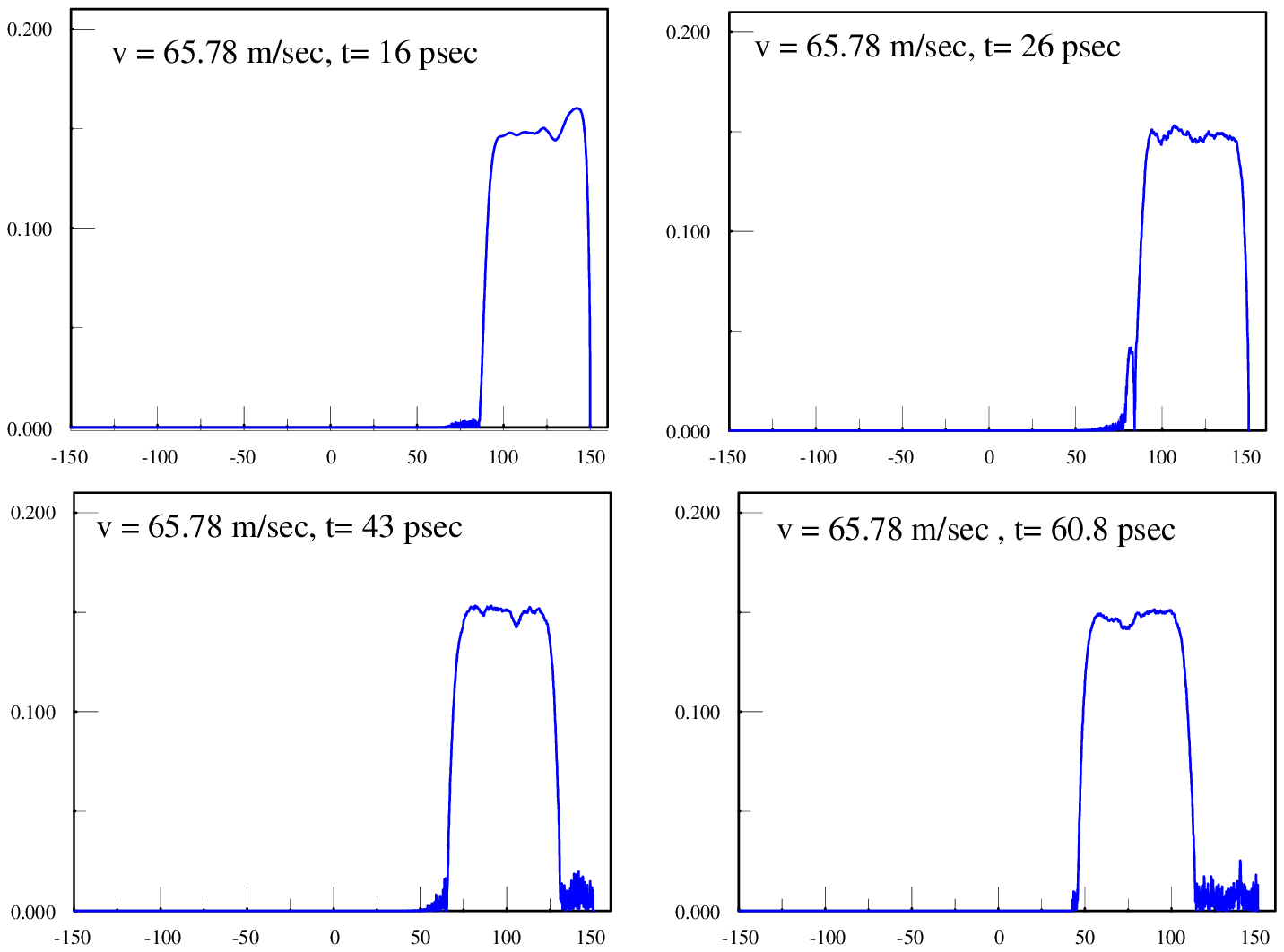}
\caption{\sl Time snapshots of $|\Psi|$ as a function 
of distance in units of $\AA$ for $v~=~65.78\frac{m}{sec}$, classical case}
\label{fig5}
\end{figure}

\begin{figure}
\epsffile{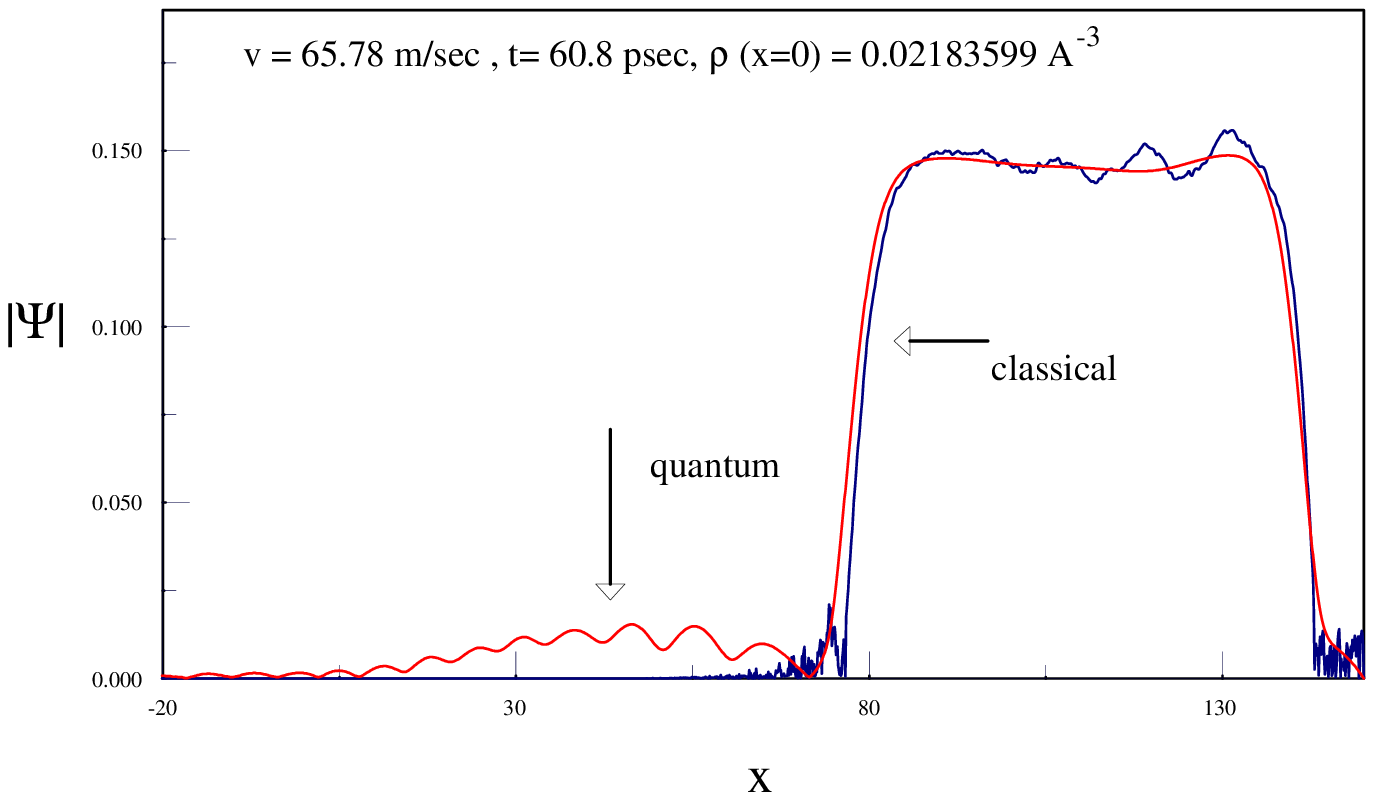}
\caption{\sl Quantum versus classical scattering for 
$\ds v~=~65.78\frac{m}{sec}$. $|\Psi|$ as a function 
of distance in units of $\AA$}
\label{fig6}
\end{figure}

\begin{figure}
\epsffile{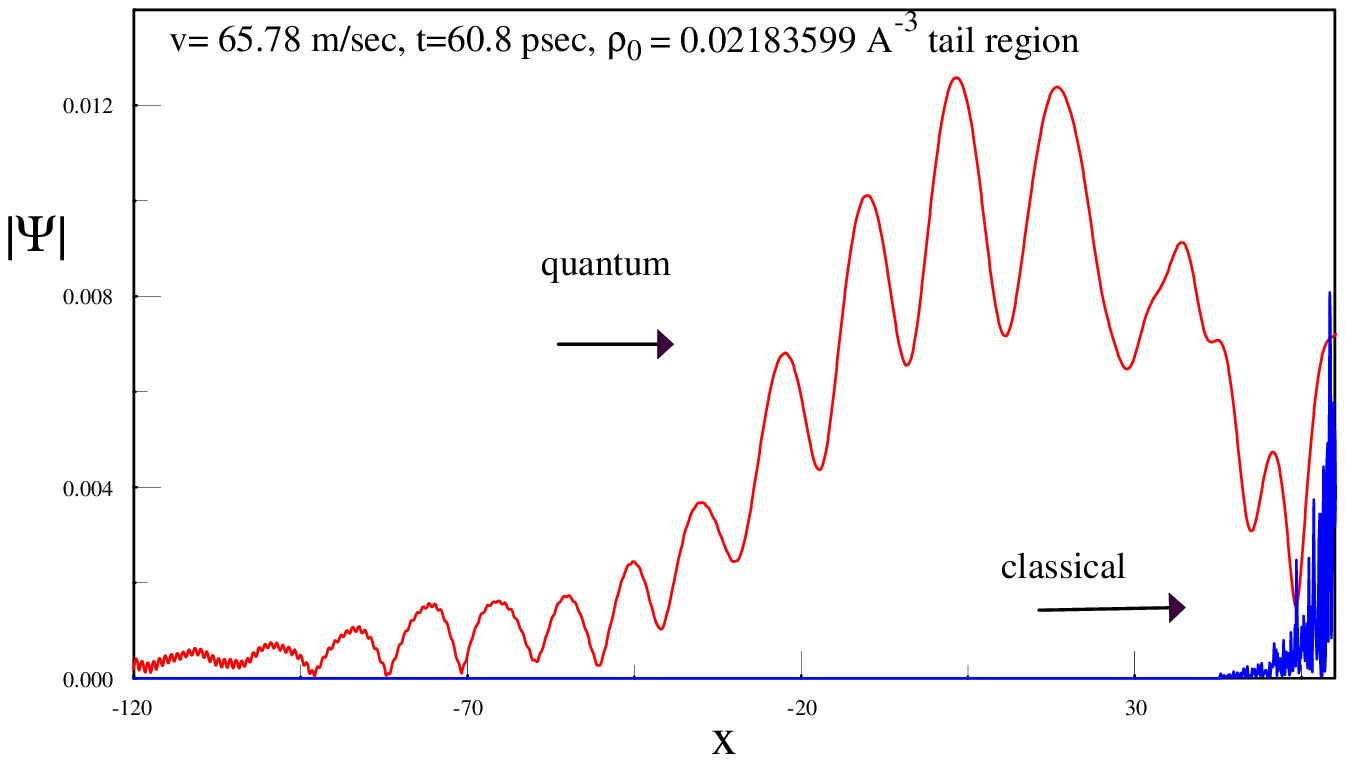}
\caption{\sl Quantum versus classical scattering for 
$\ds v~=~65.78\frac{m}{sec}$, enlargement of tail region. 
$|\Psi|$ as a function of distance in units of $\AA$}
\label{fig7}
\end{figure}

\begin{figure}
\epsffile{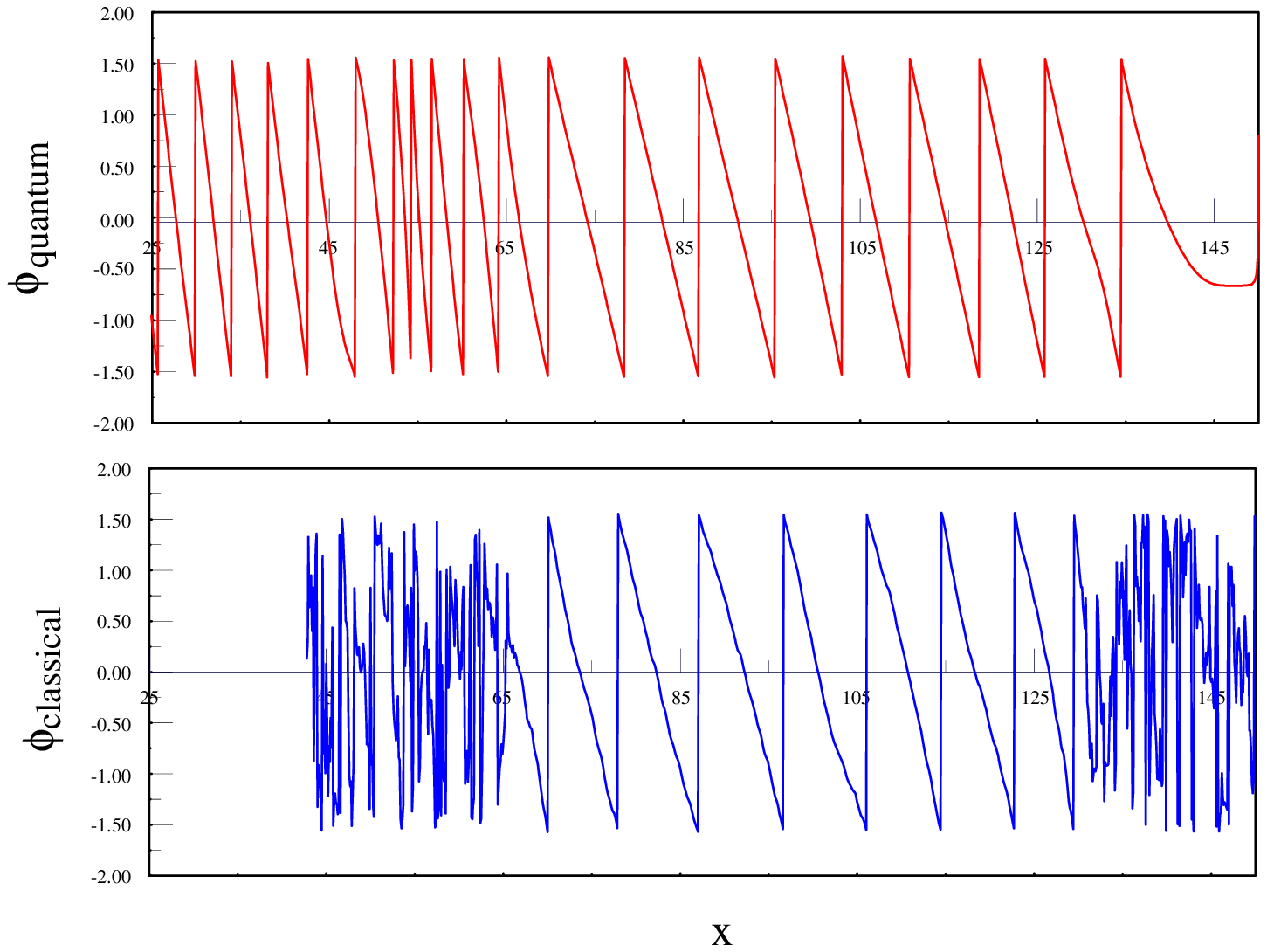}
\caption{\sl Wave function phase $\phi~=~tan^{-1}[\frac{Im(\Psi)}{Re(\Psi)}]$,
for the quantum and classical cases, as a function of distance 
in units of $\AA$, $\ds v~=~65.78\frac{m}{sec}$}
\label{fig8}
\end{figure}

Figure 6 shows classical and quantum scattering outcomes together 
at $\ds t~=~60.8~psec$. Clearly, the
quantum scattering resists distortions in the bulk and gets rid of
the energy in the form of coherent wave trains, whereas the classical
 scattering that lacks the quantum potential curvature-like term, renders
the packet softer to deformations.
The detail of the backwards region can be seen in figure 7.

The coherence of the tail region is visualized in figure 8.
In this picture we display the phase of the wave function for the scattering
case of figure 6. The classical scattering phase at the bottom is coherent
 inside the drop corresponding to a travelling packet of fixed velocity, 
and incoherent or even random at the surface. The quantum phase is coherent
both inside and outside (only part of the outside region is shown). 
The wavelength outside the bulk of the drop in the upper graph
, for $\ds x~<~70~\AA$ 
is smaller than the wavelength in the bulk. The multiple peak 
tail recedes faster than the drop.

\newpage
\begin{figure}
\epsffile{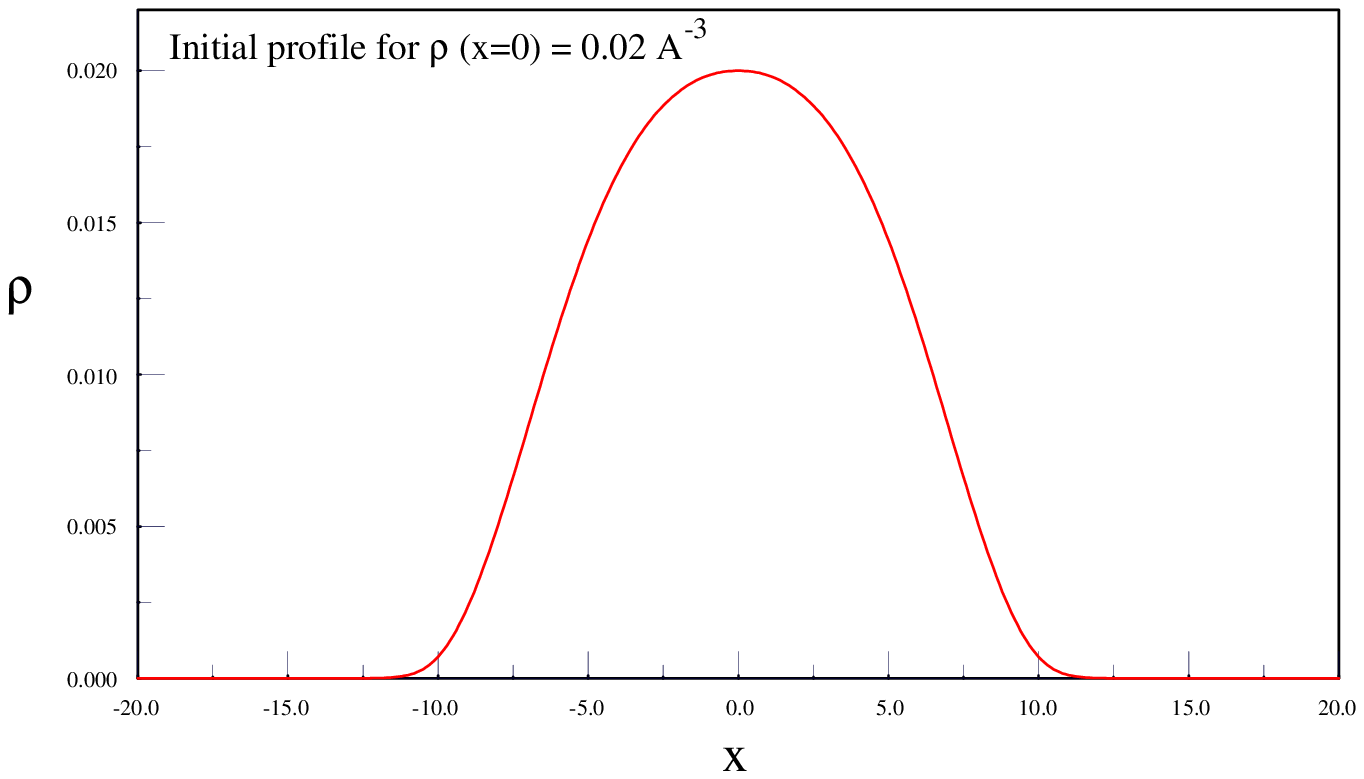}
\caption{\sl Density profile of a Helium drop
in $\AA^{-3}$ of for $\rho(x=0)~=~0.02~\AA^{-3}$ as a function
of distance in $\AA$}
\label{fig9}
\end{figure}

\section{\sl Conclusions}

\begin{figure}
\epsffile{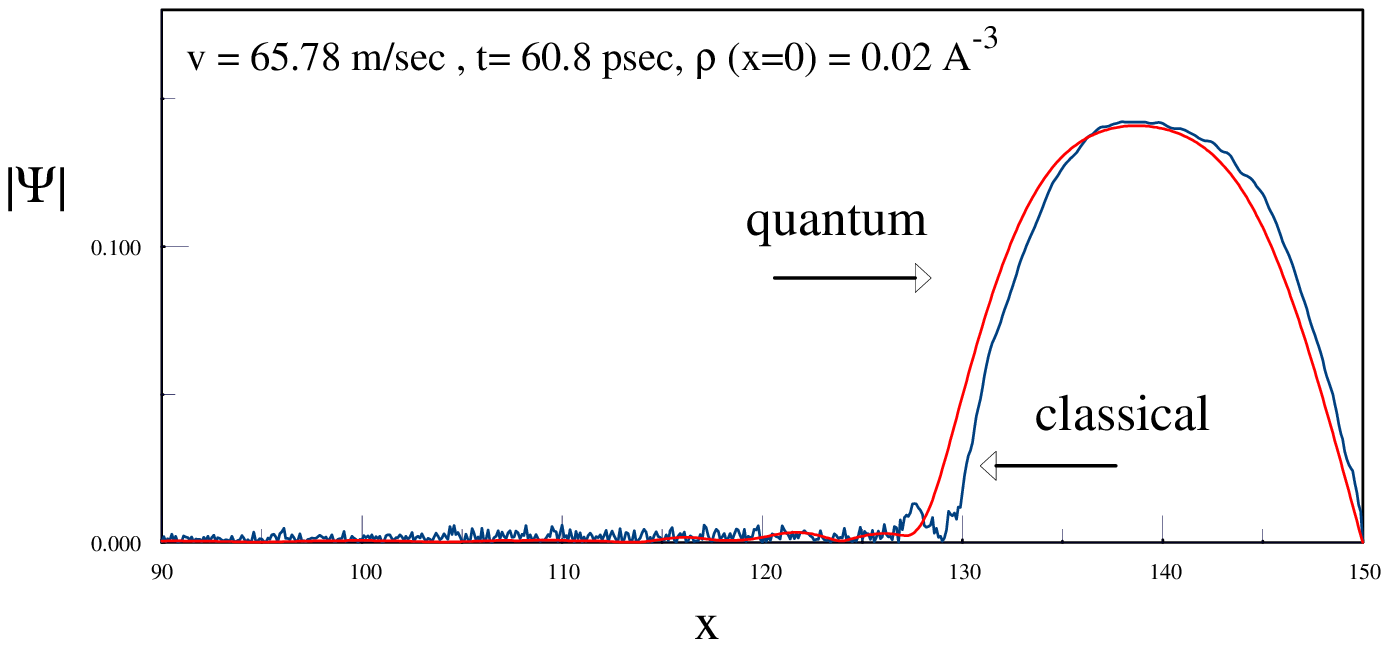}
\caption{\sl Quantum versus classical scattering for 
$\ds v~=~65.78\frac{m}{sec}$, $\rho(x=0)~=~0.02~\AA^{-3}$ . 
$|\Psi|$ as a function  of distance in units of $\AA$}
\label{fig10}
\end{figure}

\begin{figure}
\epsffile{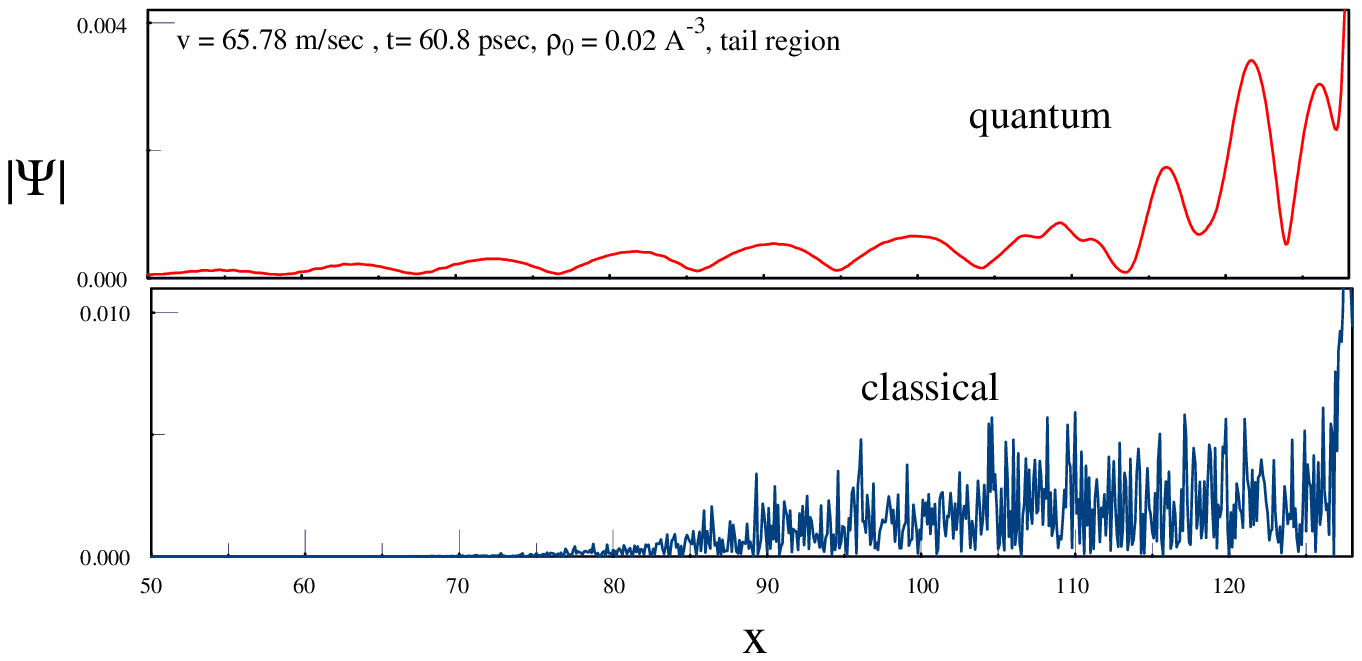}
\caption{\sl Quantum versus classical scattering for 
$\ds v~=~65.78\frac{m}{sec}$, $\rho(x=0)~=~0.02~\AA^{-3}$ , 
enlargement of tail region. 
$|\Psi|$ as a function of distance in units of $\AA$}
\label{fig11}
\end{figure}

We have confirmed numerically that the effect of {\it diffraction
of wave packets in space and time} does express itself in Helium$^4$
nanodrop scattering. The results for a thick drop presented in
the last section are not that clean as those of 
ref.\ci{k3}. A thick drop has a sizeable self-interaction
and tends to conserve its shape in the collision. To enhance the
effect, we considered also smaller size drops as the one depicted
 in figure 9. Here we used $\rho(x=0)~=~0.02~\AA^{-3}$
The number of particles per unit area is now $N~=~0.26~\AA^{-2}$, 
and the effective extent is $\ds~X_{eff}~=~12.5~\AA$. 
A three dimensional drop of this type contains around 40 atoms.

Figures 10 and 11 depict the classical and quantum 
scattering results for the 
profile of figure 9 and a velocity of $\ds v~=~65.78~\frac{m}{sec}$
at $t~=~60.8~psec$.
The asymptotic behavior of the quantum event is much cleaner
than that of the thicker drop depicted in the previous section.
In summary, it appears that it could be possible to observe
the diffraction phenomenon described in \ci{k1}-\ci{k6} with
Helium$^4$ nanodrops by colliding them with a hard surface at
high enough speed.

\end{document}